\documentclass[12pt]{article}
\usepackage{amsmath}
\usepackage{amssymb}
\usepackage{amstext}
\usepackage{amscd}
\usepackage{amsfonts}
\usepackage{appendix}
\DeclareMathAlphabet{\EuFrak}{U}{euf}{m}{n}
\DeclareMathAlphabet{\EuScript}{U}{eus}{m}{n}

\title{{\bf General Solution of a Fractional Diffusion-Advection
Equation for Solar Cosmic-Ray Transport}}
\author{M.C.Rocca$^{1,2}$,A.R.Plastino$^{3}$,\\A.Plastino$^{1,2}$,A.L.De
Paoli$^{1,2}$\\\\
\small{$^1$Departamento de F\'{\i}sica, Fac. de Ciencias Exactas},\\
\small{Universidad Nacional de La Plata}\\
\small{C.C. 67 (1900) La Plata, Argentina}\\
\small{$^2$IFLP-CCT-CONICET-C.C 727 (1900) La Plata. Argentina}\\
\small{$^3$ CeBio y Secretaria de Investigacion,}\\
\small{Univ. Nac. del Noroeste de la Prov. de Bs. As.,}\\
\small{ UNNOBA and CONICET, R. Saenz Pe\~{n}a 456, Junin, Argentina}}

\date{\today}

\begin{document}

\maketitle

\begin{abstract}

In this effort  we exactly solve the fractional
diffusion-advection equation for solar cosmic-ray transport
proposed in \cite{LE2014} and give its {\it general solution}
in terms of hypergeometric distributions. Also, we regain  all the
results and approximations given in \cite{LE2014} as {\it
particular cases} of our general solution.

Keywords:

\end{abstract}

\newpage

\renewcommand{\theequation}{\arabic{section}.\arabic{equation}}

\section{Introduction}

There is a considerable body of evidence, from data
collected by spacecrafts like {\it Ulysses} and {\it Vayager 2},
indicating that the transport of energetic particles in the
turbulent heliospheric medium is superdiffusive
\cite{PZ2007,PZ2009}. Considerable effort has been devoted in
recent years to the development of superdiffusive models for the
transport of electrons and protons in the heliosphere
\cite{SS2011,TZ2011,ZP2013}. This kind of transport regime
exhibits a power-law growth of the mean square displacement of the
diffusing particles, $\langle \Delta x^2 \rangle \propto
t^{\alpha}$, with $\alpha > 1$ (see, for instance, \cite{SZ97}).
The special case $\alpha = 2$ is called ballistic transport. The
limit case $\alpha \to 1$ corresponds to normal diffusion,
described by the well-known Gaussian propagator. The energetic
particles detected by the aforementioned probes are usually
associated with violent solar events like solar flares. These
particles diffuse in the solar wind, which is a turbulent
environment than can be assumed statistically homogeneous at large
enough distances from the sun \cite{PZ2007}. This implies that the
propagator $P(x,x',t,t')$, describing the probability of finding
at the space time location $(x,t)$ a particle that has been
injected at $(x',t')$, depends solely on the differences $x-x'$
and $t-t'$. In the superdiffusive regime the propagator
$P(x,x',t,t')$ is not Gaussian, and exhibits power-law tails. It
arises as solution a non local diffusive process governed by an
integral equation that can be recast under the guise of a
diffusion equation where the well-known Laplacian term is replaced
by a term involving fractional derivatives \cite{C95}. Diffusion
equations with fractional derivatives have attracted considerable
attention recently (see \cite{LTRLGS2014,
LSSEL2009,RLEL2007,stern,LMASL2005} and references therein) and have
lots of potential applications \cite{MK2000,P2013}. In particular,
the observed distributions of solar cosmic ray particles are often
consistent with power-law tails, suggesting that a superdiffusive
process is at work.

A proper understanding of the transport of energetic
particles in space is a vital ingredient for the analysis of
various important phenomena, such as the propagation of particles
from the Sun to our planet or, more generally, the acceleration
and transport of cosmic rays. The superdiffusion of particles in
interplanetary turbulent environments is often modelled using
asymptotic expressions for the pertinent non-Gaussian propagator,
which have a limited range of validity. A first step towards a
more accurate analytical treatment of this problem was recently
provided by Litvinenko and Effenberger (LE) in \cite{LE2014}. LE
considered solutions of a fractional diffusion-advection equation
describing the diffusion of particles emitted at a shock front
that propagates at a constant upstream speed $V_{sh}$ in the solar
wind rest frame. The shock front is assumed to be planar, leading
to an effectively one-dimensional problem. Each physical quantity
depends only on the time $t$ and on the spatial coordinate $x$
measured along an axis perpendicular to the shock front. In the
present contribution we re-visit the fractional
diffusion-advection equation previously studied by LE, providing a
closed analytical solution.

\section{Formulation of the Problem}

\setcounter{equation}{0}

The authors of  \cite{LE2014} advanced  the equation
\begin{equation}
\label{ep2.1} 
\frac {\partial f} {\partial t}=\kappa\frac
{{\partial}^{\alpha} f} {\partial |x|^{\alpha}}+a\frac {\partial
f} {\partial x}+\delta(x),
\end{equation}
where $t>0$ and $f(x,t)$ is the distribution function for solar
cosmic-rays transport. Here the fractional spatial derivative is
defined as
\begin{equation}
\label{ep2.2} 
\frac {{\partial}^{\alpha} f} {\partial
|x|^{\alpha}}= \frac {1} {\pi}\sin\left(\frac {\pi\alpha}
{2}\right)\Gamma(\alpha+1) \int\limits_0^{\infty}\frac
{f(x+\xi)-2f(x)+f(x-\xi)} {{\xi}^{\alpha+1}}\;d\xi.
\end{equation}
(See \cite{LE2014} and references therein).

To solve this equation the authors use the Green function governed
by the equation:
\begin{equation}
\label{ep2.3} 
\frac {\partial {\cal G}} {\partial t}=\kappa\frac
{{\partial}^{\alpha} {\cal G}} {\partial
|x|^{\alpha}}+\delta(x)\delta(t).
\end{equation}
With this Green function, the solution of (\ref{ep2.1}) can be
expressed as
\begin{equation}
\label{ep2.4} 
f(x,t)=\int\limits_0^t {\cal G}(x+at^{'},
t^{'})\;dt^{'}.
\end{equation}
In this work we obtain the solutions of Eqs. (\ref{ep2.1}) and
(\ref{ep2.3})  using  distributions as main tools
 \cite{guelfand1}. Also, we re-obtain all results and approximations
obtained in \cite{LE2014}, but  as particular cases of our general
solutions of (\ref{ep2.1}) and (\ref{ep2.2}).

For our task we use, as a first step, the solution obtained in
\cite{LE2014} for the Green function through the use of the Fourier
Transform given by
\begin{equation}
\label{ep2.5} 
\hat{{\cal G}}(k,t)=\frac {1} {2\pi}
\int\limits_{-\infty}^{\infty}{\cal G}(x,t) e^{-ikx}\;dx,
\end{equation}
from which we obtain for $\hat{{\cal G}}$:
\begin{equation}
\label{ep2.6} 
\hat{{\cal G}}(k,t)=-\kappa|k|^{\alpha}\hat{{\cal
G}}(k,t)+\frac {1} {2\pi}\delta(t),
\end{equation}
whose solution is
\begin{equation}
\label{ep2.7} 
\hat{{\cal G}}(k,t)=\frac {H(t)} {2\pi} e^{-\kappa
|k|^{\alpha}t},
\end{equation}
where $H(t)$ is the Heaviside's step function.

\section{General Solution of the Equations}

\setcounter{equation}{0}

From (\ref{ep2.7}) we have for $\hat{{\cal G}}$
\begin{equation}
\label{ep3.1} 
\hat{{\cal G}}(k,t)=\frac {H(t)} {2\pi} e^{-\kappa
|k|^{\alpha}t}= \frac {H(t)} {2\pi}
\sum\limits_{n=0}^{\infty}\frac {(-1)^n{\kappa}^nk^{\alpha n} t^n}
{n!},
\end{equation}
and, invoking  the inverse Fourier transform
\[{\cal G}(x,t)=\frac {H(t)} {2\pi}
\int\limits_{-\infty}^{\infty}
e^{-\kappa |k|^{\alpha}t}e^{ikx}\;dk=\]
\begin{equation}
\label{ep3.2} 
\frac {H(t)} {2\pi} \sum\limits_{n=0}^{\infty}\frac
{(-1)^n{\kappa}^n t^n} {n!} \left[\int\limits_0^{\infty}k^{\alpha
n}e^{ikx}\;dx + \int\limits_0^{\infty}k^{\alpha
n}e^{-ikx}\;dx\right].
\end{equation}
Fortunately, we can find in the classical book of \cite{guelfand1}
 the results for the two integrals of (\ref{ep3.2}). We
obtain
\begin{equation}
\label{ep3.3} 
{\cal G}(x,t)=\frac {H(t)} {2\pi}
\sum\limits_{n=0}^{\infty}\frac {(-1)^n{\kappa}^n t^n} {n!}
\Gamma(\alpha n+1) \left[\frac {e^{i\frac {\pi} {2}(\alpha n+1)}}
{(x+i0)^{\alpha n + 1}} + \frac {e^{-i\frac {\pi} {2}(\alpha
n+1)}} {(x-i0)^{\alpha n + 1}} \right].
\end{equation}
Using now (\ref{ep2.4}) we have for $f$
\[f(x,t)=\int\limits_0^t
{\cal G}(x+at^{'}, t^{'})\;dt^{'},\] so that one can write
\[f(x,t)=\frac {1} {2\pi}
\sum\limits_{n=0}^{\infty}\frac {(-1)^n{\kappa}^n} {n!}
\Gamma(\alpha n+1)\times\]
\begin{equation}
\label{ep3.4}
 \int\limits_0^t \left[\frac {e^{i\frac {\pi}
{2}(\alpha n+1)}} {(x+at^{'}+i0)^{\alpha n + 1}} + \frac
{e^{-i\frac {\pi} {2}(\alpha n+1)}} {(x+at^{'}-i0)^{\alpha n + 1}}
\right]t^{'n}\;dt^{'}.
\end{equation}
According to Eq. (\ref{a1}) of the  Appendix, where $t>0$, we now
obtain for $f$, invoking hypergeometric functions
$F(\alpha n+1,2;3;z)$ and Beta functions ${\cal B}(1,n+1)$,
\[f(x,t)=\frac {1} {2\pi}\sum\limits_{n=0}^{\infty}
 \frac {(-1)^n{\kappa}^nt^{n+1}} {n!}\Gamma(\alpha n+1)
{\cal B}(1,n+1)\times\]
\[\left[\frac {e^{i\frac {\pi} {2}(\alpha n + 1)}}
 {(x+i0)^{\alpha n + 1}}
F\left(\alpha n+1,n+1;n+2;-\frac {at} {x+i0}\right)+\right.\]
\begin{equation}
\label{ep3.5}
\left.\frac {e^{-i\frac {\pi} {2}(\alpha n + 1)}}
 {(x-i0)^{\alpha n + 1}}
F\left(\alpha n+1,n+1;n+2;-\frac {at} {x-i0}\right)\right].
\end{equation}
This is the general solution of Eq.(\ref{ep2.1}) for the initial
condition $f(x,0)=0$.

In the next section we will see that  all results and
approximations obtained in \cite{LE2014} can be regarded as
particular cases of the general solution (\ref{ep3.5}).

\section{Weak Diffusion Approximation}

\setcounter{equation}{0}

Following LE, we shall now consider a weak diffusion
approximation. Within this approximation we can treat $\kappa$ as
a small parameter and develop $f$ up to order one \cite{LE2014}.
Thus, we can write:
\begin{equation}
\label{ep4.1} 
f(x,t)=f_0(x,t)+f_1(x,t),
\end{equation}
where i) $f_0$ is independent of $\kappa$ and  ii) in $f_1$ the
corresponding power of $\kappa$ is unity.

Eq. (\ref{ep3.5}) entails that  we have, for $f_0$ ($n=0$ in
(\ref{ep3.5})),
\[f_0(x,t)=\frac {it} {2\pi}\left[
(x+i0)^{-1}F\left(1,1;2;-\frac {at} {x+i0}\right)\right.-\]
\begin{equation}
\label{ep4.2} 
\left.(x-i0)^{-1}F\left(1,1;2;-\frac {at}
{x-i0}\right)\right].
\end{equation}
Recourse to the celebrated Tables  of \cite{gra1} allows us to
write
\begin{equation}
\label{ep4.3} 
F(1,1;2;-z)=\frac {1} {z}\ln(1+z),
\end{equation}
and we obtain for $f_0$
\begin{equation}
\label{ep4.4} 
f_0(x,t)=\frac {1} {a}[H(-x)-H(-x-at)=\frac {1} {2a}
[Sgn(x+at)-Sgn(x)].
\end{equation}
When we take $n=1$ in (\ref{ep3.5}), $f_1$ is defined as
\[f_1(x,t)=-\frac {i\kappa t^2} {4\pi}\Gamma(\alpha+1)\left[
\frac {e^{i\frac {\pi} {2}}} {(x+i0)^{\alpha+1}}
F\left(\alpha+1,2;3;-\frac {at} {x+i0}\right)\right.\]
\begin{equation}
\label{ep4.5} 
+\left.\frac {e^{-i\frac {\pi} {2}}}
{(x-i0)^{\alpha+1}} F\left(\alpha+1,2;3;-\frac {at}
{x-i0}\right)\right].
\end{equation}
Now, from (\ref{a1}) of Appendix we have, for the hypergeometric
function,
\[F(\alpha+1,2;3;z)=\frac {2} {\alpha(\alpha-1)z^2}
\left[1+\frac {\alpha z-1} {(1-z)^{\alpha}}\right],\] so that,
using this result, $f_1$ adopts the form
\[f_1(x,t)=\frac {i\kappa\Gamma(\alpha-1)} {2\pi a^2}
\left\{(x+\alpha at)
\left[\frac {e^{i\frac {\pi} {2}\alpha}} {(x+at+i0)^{\alpha}}-
\frac {e^{-i\frac {\pi} {2}\alpha}} {(x+at-i0)^{\alpha}}\right]+
\right.\]
\begin{equation}
\label{ep4.6} 
\left.\frac {e^{-i\frac {\pi} {2}\alpha}}
{(x-i0)^{\alpha-1}}- \frac {e^{i\frac {\pi} {2}\alpha}}
{(x+i0)^{\alpha-1}}\right\}.
\end{equation}
Using at this point  (\ref{ep4.1}), (\ref{ep4.4}), and
(\ref{ep4.6}),  the final result for $f$,  up to first order in
$\kappa$, reads, invoking the sign function  $Sgn(x)$,
\[f(x,t)=\frac {1} {2a}[Sgn(x+at)-Sgn(x)]+\]
\[\frac {i\kappa\Gamma(\alpha-1)} {2\pi a^2}
\left\{(x+\alpha at)
\left[\frac {e^{i\frac {\pi} {2}\alpha}} {(x+at+i0)^{\alpha}}-
\frac {e^{-i\frac {\pi} {2}\alpha}} {(x+at-i0)^{\alpha}}\right]+
\right.\]
\begin{equation}
\label{ep4.7} 
\left.\frac {e^{-i\frac {\pi} {2}\alpha}}
{(x-i0)^{\alpha-1}}- \frac {e^{i\frac {\pi} {2}\alpha}}
{(x+i0)^{\alpha-1}}\right\}.
\end{equation}
From this expression for $f$, we will obtain all approximate
results reported  in \cite{LE2014}. Thus, for $x>0$ (\ref{ep4.7})
becomes
\begin{equation}
\label{ep4.8} 
f(x,t)=\frac {\kappa\sin(\frac {\pi\alpha}
{2})\Gamma(\alpha-1)} {\pi a^2} \left[\frac {1} {x^{\alpha-1}}-
\frac {x+\alpha at} {(x+at)^{\alpha}}\right].
\end{equation}
We have to distinguish two limiting cases. The first one is the
asymptotic situation  $x>>at$.  In this case,
\begin{equation}
\label{ep4.9} 
f(x,t)=\frac {1} {2\pi} \sin(\frac {\pi\alpha} {2})
\Gamma(\alpha+1)\frac {\kappa t^2} {x^{\alpha+1}}.
\end{equation}
The second case limiting case is  $0<x<<at$. The corresponding
expression for $f$ becomes
\begin{equation}
\label{ep4.10} 
f(x,t)=\frac {1} {\pi} \sin(\frac {\pi\alpha} {2})
\Gamma(\alpha-1)\frac {\kappa} {a^2}x^{1-\alpha}.
\end{equation}\vskip 4mm

We consider signs now. When $x+at<0$, from (\ref{ep4.7}) we have
\begin{equation}
\label{ep4.11}
f(x,t)=\frac {\kappa\sin(\frac {\pi\alpha} {2})\Gamma(\alpha-1)}
{\pi a^2} \left[\frac {1} {|x|^{\alpha-1}}+
\frac {x+\alpha at} {|x+at|^{\alpha}}\right]
\end{equation}
Again, two special cases must be considered. One is for $x<<-at$
for which
\begin{equation}
\label{ep4.12} 
f(x,t)=\frac {1} {2\pi} \sin(\frac {\pi\alpha} {2})
\Gamma(\alpha+1)\frac {\kappa t^2} {|x|^{\alpha+1}}.
\end{equation}
The other special  situation  is  $x<0$, $x+at>0$, $x>>-at$. Here,

\begin{equation}
\label{ep4.13} 
f(x,t)=\frac {1} {a}+\frac {1} {\pi} \sin(\frac
{\pi\alpha} {2}) \Gamma(\alpha-1)\frac {\kappa}
{a^2}|x|^{1-\alpha}.
\end{equation}
At this stage, we have re-obtained all approximations given in
\cite{LE2014}, but using  a more general procedure. More
specifically, all approximations have been obtained from only one
relation: Eq. (\ref{ep4.7}), which, in turn, is deduced from our
general formula (\ref{ep3.5}).

\section{Change of Variables}

\setcounter{equation}{0}

We assume that in the solar wind rest frame the particles'
transport is described by the fractional-diffusion equation with
no advection term (that is, with $a=0$ in (\ref{ep2.1})). The
shock front (which started at $x_0 = -V_{sh} t_0$, moves with
constant speed $V_{sh}$, and is regarded as highly localized in
the $x$-coordinate) constitutes the source of the particles.
Consequently, we have a fractional-diffusion equation with a
uniformly moving Dirac's delta source of the form $\delta(x -
V_{sh} t)$. In order to have a stationary delta source we need to
perform an appropriate change of coordinates, re-casting our
problem in a reference frame where the shock front is stationary.
We also change the origin of time so that the source starts being
active at $t=0$. In this new reference frame the transport
equation has an advection term with advection velocity $a=
V_{sh}$, and a stationary source $\delta (0)$ that starts at
$t=0$. After solving the diffusion-advection equation in this new
frame (which is what we have done in the previous sections) we
re-express the solution in terms of the original coordinates
associated with the solar wind rest frame. This last step is
succinctly described by the three correspondences $a\rightarrow
v_{sh}$, $t\rightarrow t+t_0$, and $x\rightarrow x-v_{sh}t$, after
which Eq. (\ref{ep3.5}) adopts the appearance
\[f(x,t)=\frac {1} {2\pi}\sum\limits_{n=0}^{\infty}
 \frac {(-1)^n{\kappa}^n(t+t_0)^{n+1}} {n!}\Gamma(\alpha n+1)
{\cal B}(1,n+1)\times\]
\[\left[\frac {e^{i\frac {\pi} {2}(\alpha n + 1)}}
{(x-v_{sh}t+i0)^{\alpha n + 1}}
F\left(\alpha n+1,n+1;n+2;-\frac {v_{sh}(t+t_0)} {x-v_{sh}t+i0}\right)+\right.\]
\begin{equation}
\label{ep5.1} 
\left.\frac {e^{-i\frac {\pi} {2}(\alpha n + 1)}}
{(x-v_{sh}t-i0)^{\alpha n + 1}} F\left(\alpha n+1,n+1;n+2;-\frac
{v_{sh}(t+t_0)} {x-v_{sh}t-i0}\right)\right].
\end{equation}
Accordingly, in  the weak diffusion approximation (\ref{ep4.7}) we
have
\[f(x,t)=\frac {1} {2v_{sh}}
[Sgn(x+v_{sh}t_0)-Sgn(x-v_{sh}t)]+\]
\[\frac {i\kappa\Gamma(\alpha-1)} {2\pi V_{sh}^2}
\left\{(x+(\alpha-1) v_{sh}t+v_{sh}t_0)
\left[\frac {e^{i\frac {\pi} {2}\alpha}} {(x+v_{sh}t_0+i0)^{\alpha}}-
\right.\right.-\]
\begin{equation}
\label{ep5.2}
\left.\left.\frac {e^{-i\frac {\pi} {2}\alpha}}
{(x+v_{sh}t_0-i0)^{\alpha}}\right]
+\frac {e^{-i\frac {\pi} {2}\alpha}} {(x-v_{sh}t-i0)^{\alpha}}-
\frac {e^{i\frac {\pi} {2}\alpha}} {(x-v_{sh}t+i0)^{\alpha}}\right\}
\end{equation}

\section{Conclusions}

 We have provided an explicit analytical solution for and
 advection-diffusion equation involving fractional derivatives in
 the diffusion term. This equation governs the diffusion of
 particles in the solar wind injected at the front of a shock
 that travels at a constant upstream speed $V_{sh}$ in the solar
 wind rest frame. The shock is assumed to have a planar front,
 leading to a problem with an effective one dimensional geometry,
 where all the relevant quantities depend solely on time and on
 the coordinate $x$ measured along an axis perpendicular to the
 front. 

 We obtained the exact solution of the above mentioned
 equation in the $x$-configuration space (besides the
 associated formal solution in the $k$-space related
 to the previous one via a Fourier transform). Our solution
 allow us to obtain in a unified and systematic way all the
 relevant approximations that were previously discussed
 by LE, each one in a separated way.

\newpage

\renewcommand{\thesection}{\Alph{section}}

\renewcommand{\theequation}{\Alph{section}.\arabic{equation}}

\setcounter{section}{1}

\section*{Appendix:Some properties of Hypergeometric Function}

\setcounter{equation}{0}

Using data from \cite{gra2} we have
\[\int\limits_0^t\frac {t^{'n}} {(x+at^{'}\pm i0)^{\alpha n +1}}\;dt^{'}=
\frac {t^{n+1}} {(x\pm i0)^{\alpha n+1}}B(1,n+1)\times\]
\begin{equation}
\label{a1} 
F\left(\alpha n +1, n+1,n+2;-\frac {at} {x\pm
i0}\right).
\end{equation}
Now we appeal to the transformation formula given in \cite{gra3}
for the hypergeometric function
\[F(\alpha+1,2;3;z)=\frac {2\Gamma(1-\alpha)} {\Gamma(2-\alpha)}
(-1)^{\alpha+1}z^{-\alpha-1}
F\left(\alpha+1,\alpha-1;\alpha;\frac {1} {z}\right)+\]
\begin{equation}
\label{a2} 
\frac {2\Gamma(\alpha-1)} {\Gamma(\alpha+1)} z^{-2}
F\left(2,0;2-\alpha;\frac {1} {z}\right),
\end{equation}
with
\begin{equation}
\label{a3} 
F(a,0;c;z)=1.
\end{equation}
In these circumstance we obtain
\[F(\alpha+1,2;3;z)=\frac {2\Gamma(1-\alpha)} {\Gamma(2-\alpha)}
(-1)^{\alpha+1}z^{-\alpha-1}
F\left(\alpha+1,\alpha-1;\alpha;\frac {1} {z}\right)+\]
\begin{equation}
\label{a4} 
\frac {2\Gamma(\alpha-1)} {\Gamma(\alpha+1)}. z^{-2}
\end{equation}
Now we invoke  the transformation formula \cite{gra4}
\begin{equation}
\label{a5} 
F\left(\alpha+1,\alpha-1;\alpha;\frac {1} {z}\right)=
\left(1-\frac {1} {z}\right)^{-\alpha}F\left(-1,1;\alpha;\frac {1}
{z}\right),
\end{equation}
or
\begin{equation}
\label{a6} 
F\left(\alpha+1,\alpha-1;\alpha;\frac {1} {z}\right)=
\frac {z^{\alpha}} {(z-1)^{\alpha}}\left(\frac {\alpha z-1}
{\alpha z}\right).
\end{equation}
At this stage,  we have, finally,
\begin{equation} 
\label{a7}
F(\alpha+1,2;3;z)=\frac {2} {\alpha(\alpha-1)z^2} \left[1+\frac
{\alpha z-1} {(1-z)^{\alpha}}\right].
\end{equation}

\end{document}